
\nopagenumbers
\magnification=\magstep1
\parskip 0pt
\parindent 15pt
\baselineskip 16pt
\hsize 5.53 truein
\vsize 8.5 truein

\font\titolo = cmbx10 scaled \magstep2
\font\autori = cmsl10 scaled \magstep2
\font\abstract = cmr9

\def\CC {{\hbox{\tenrm C\kern-.45em{\vrule height.67em width0.08em
        depth-.04em \hskip.45em}}}}
\def\DD {{\hbox{\tenrm {I\kern-.18em{D}}\kern-.36em {\vrule
        height.62em width0.08em depth-.04em\hskip.36em}}}}

\def\RR {{\hbox{\tenrm I\kern-.17em{R}}}}
\def\BB {{\hbox{\tenrm I\kern-.17em{B}}}}
\def\ZZ {{\hbox{\tenrm Z\kern-.31em{Z}}}}
\def\ZB {{\hbox{\tenrm Z\kern-.31em{Z}$_2$}}}
\def\KK {{\hbox{\tenrm I\kern-.17em{K}}}}

\def\olf {{\hbox{\tensy O\kern-.68em{O}$_f$}}}
\def\olfa {{\hbox{\tensy O\kern-.68em{O}$_{f_1}$}}}
\def\olh {{\hbox{\tensy O\kern-.68em{O}$_h$}}}
\def\olx {{\hbox{\tensy O\kern-.68em{O}$_x$}}}

\def\gsh {{\hbox{\tensy M\kern-1.14em{M}}}}
\def\id {{\hbox{\tenrm I\kern-.19em{I}}}}
\def\jj {{\hbox{\tenrm J\kern-.35em{J}}}}

\def\qq {{\hbox{\tentt [\kern-0.07em{]}}}}

\def\cc {{\hbox{\rm C\kern-.45em{\vrule height.67em width.08em depth-.04em
\hskip.45em}}}}
\def\ii {\hbox{\rm I\kern-.19em{I}}}
\def\zz {\hbox{\rm Z\kern-.35em{Z}}}
\def\rr {\hbox{\rm I\kern-.17em{R}}}
\def\mm{{\hbox{\tensy M\kern-1.14em{M}}}}
\def\hh {{\hbox{\tensy H\kern-.78em{H}}}}
\def\jj {{\hbox{\tenrm J\kern-.35em{J}}}}

\hrule height 0pt
\vskip 0.5 truein
\line{\hfil POLFIS-TH05/92}
\vfill
\centerline{\titolo MULTICRITICAL POINTS AND}
\vskip .05truein
\centerline{\titolo REENTRANT PHENOMENON}
\vskip .05truein
\centerline{\titolo IN THE BEG MODEL}
\vskip .15truein
\centerline{\autori Carla Buzano and Alessandro Pelizzola}
\vskip .15truein
\centerline{\abstract Dipartimento di Fisica, Politecnico di Torino,
10129 Torino, Italy}
\vskip .35truein
\moveright 1.5 truein \vbox{\hrule width 2.9 truein height 1pt}
\vskip .35truein
\centerline{\bf ABSTRACT}
\vskip .05truein {\abstract The Blume - Emery - Griffiths model is
investigated by
use of the cluster variation method in the pair approximation. We determine
the regions of the phase space where reentrant phenomenon takes place. Two
regions are found, depending on the sign of the reduced quadrupole - quadrupole
coupling strength $\xi$. For negative $\xi$ we find Para-Ferro-Para and
Ferro-Para-Ferro-Para transition sequences; for positive $\xi$, a
Para$_-$-Ferro-Para$_+$ sequence. Order parameters, correlation functions
and specific heat are given in some typical cases. By-products of this work
are the equations for the critical and tricritical lines.} \par
\vskip .35 truein
\vfill
\halign{#&#\hfil\cr
Keywords: & spin-1 model, \cr
& reentrant phenomenon, \cr
& multicritical points. \cr}
\vskip .15 truein
\parindent 0pt
PACS numbers: 64.60-Cn; 75.10-Hk \par
\vskip .15 truein
Running title: Reentrant phenomenon in the BEG model.
\eject

\pageno=1
\footline{\hss\tenrm\folio$\,$\hss}
\def\makefootline{\baselineskip=50pt\line{\the\footline}}
\parindent 15pt

{\bf 1. Introduction} \par
\smallskip
The Blume - Emery - Griffiths (BEG) model has originally been proposed to
explain some features of the thermodynamic behavior of He$^3$ - He$^4$
mixtures$^{[1]}$ and has subsequently been used to describe the critical
properties of other physical systems like multicomponent fluids$^{[2]}$ and
magnetic systems$^{[3]}$. \par
Since its introduction, it has been studied by many authors using a lot of
different
techniques: mean field approximation$^{[1]}$, high temperature series
expansion$^{[4]}$,
Monte Carlo methods$^{[5]}$, renormalization group$^{[6-8]}$,
effective field theories$^{[9-11]}$,
cluster variation method (CVM)$^{[12]}$. In the last few years, some
authors$^{[8-12]}$ pointed out the occurrence of a reentrance in the
ferromagnetic-paramagnetic transition surface. In this paper, we use the
cluster variation method$^{[13]}$ (in the new formulation by An$^{[14]}$,
which resorts to the
M\"obius inversion) in the pair approximation, to determine the region of
the phase space of the model in which reentrance occurs. Two different
regions are found, depending on the sign of the reduced quadrupole - quadrupole
coupling strength $\xi$. In the negative $\xi$ region, the shape of the
phase diagram leads us to characterize two kinds of reentrance: the first
one is given
by a first order Ferro-Para transition followed by Para-Ferro and
Ferro-Para second order transitions; the second one is analogous except
that it does not exhibit the first order transition. In the positive $\xi$
region, a new Para$_-$-Ferro-Para$_+$ reentrant phenomenon is noticed.
In the most significant cases we give plots of order parameters,
correlation functions and specific heat. Comparison of our results with
those obtained with other techniques is made.
By-products of this work are the equations for the critical
and tricritical lines. \par
The paper is organized as follows: in Sec. 2 we write the CVM free energy
for the BEG model in pair approximation
and the equations for the order parameters and the correlation functions;
in Sec. 3 we determine the equations for the critical and tricritical
points, that will be used in Sec. 4 to determine the reentrance regions; in
Sec. 4 we also give specific heat, order parameters and correlation
functions for some typical cases of reentrant phenomenon; in Sec. 5 some
conclusions are drawn.
\par
\medskip

{\bf 2. CVM free energy and equations for the BEG model} \par
\smallskip
The BEG model has hamiltonian$^{[1]}$
$$H_{BEG} = - J \sum_{\langle i,j \rangle} S_i S_j - G \sum_{\langle i,j
\rangle}
S_i^2 S_j^2 + D \sum_i S_i^2 - B \sum_i S_i, \eqno(1)$$
where $i,j$ are lattice site labels, $S_i$ is the $z$-component of a spin 1
operator at site $i$, $\sum_{\langle i,j \rangle}$ is a sum over all
the nearest neighbors (n.n.), and $J > 0$ (ferromagnetic case).
In the pair approximation of the CVM
(corresponding to the Bethe approximation$^{[15,16]}$) one
introduces the order parameters $y_1 = \langle S_i \rangle$
($\langle \ \rangle$ denotes thermal average) and
$y_2 = \langle S_i^2 \rangle$ and the n.n. correlation functions
$y_3 = \langle S_i S_j \rangle$, $y_4 = \langle S_i S_j^2 \rangle$ and
$y_5 = \langle S_i^2 S_j^2 \rangle$ and considers the average value of
the reduced hamiltonian
$${\cal H} = {\langle H_{BEG}/N \rangle \over z J} =
dy_2 - {1 \over 2}(y_3 + \xi y_5) - b y_1, \eqno(2)$$
where $b = B/z J$, $d = D/z J$ and $\xi = G/J$ are adimensional parameters
and $z$ is the coordination number of the lattice. \par
The reduced free energy per site
$f = {\displaystyle {F/N \over z J} = {\cal H} - {k_B T \over z J}{S\over N}}$
(where $S$ is the entropy)
has been calculated in our approximation in Ref. 12, following An's
scheme$^{[14]}$, in terms of the site and pair density matrices
$$f = {\cal H} + \tau\left[(1 - z){\rm Tr}(\rho_s \ln \rho_s) +
{z \over 2}{\rm Tr}(\rho_p \ln \rho_p)\right], \eqno(3)$$
with $\tau = k_B T/z J$. $\rho_s$ and $\rho_p$ are respectively the site
and pair density matrices, that turn out to be diagonal with elements
$$\rho_{s_1} = {y_2 + y_1 \over 2}, \qquad
\rho_{s_2} = 1 - y_2, \qquad
\rho_{s_3} = {y_2 - y_1 \over 2}; \eqno(4)$$
$$\rho_{p_1} = {y_3 + 2y_4 + y_5 \over 4}, \quad
\rho_{p_9} = {y_3 - 2y_4 + y_5 \over 4},$$
$$\rho_{p_2} = \rho_{p_4} = {y_2 - y_5 + y_1 - y_4 \over 2}, \quad
\rho_{p_6} = \rho_{p_8} = {y_2 - y_5 - y_1 + y_4 \over 2},$$
$$\rho_{p_3} = \rho_{p_7} = {y_5 - y_3 \over 4}, \quad
\rho_{p_5} = 1 + y_5 - 2y_2.\eqno(5)$$
It is easy to verify that the relations
$${\rm Tr}\rho_s = {\rm Tr}\rho_p = 1, \qquad
\rho_s =
\matrix{\phantom{^{p \backslash s}} \cr
{\rm Tr} \cr
^{p \backslash s} \cr}
\rho_p \eqno(6)$$
hold (the last Tr must be intended as a partial trace over the pair degrees
of freedom, keeping fixed the site ones).
In addition, the $y_i$'s must obey the constraints given by
$$ 0 \le \rho_{s_i}, \rho_{p_i} \le 1 \eqno(7)$$
and have to be determined minimizing the free energy $f$. \par
The minimum-$f$ requirement leads to the system of nonlinear equations
$${\partial f \over \partial y_i} = 0, \qquad i = 1, \ldots 5. \eqno(8)$$
This is a system in the $y_i$'s, that can be solved explicitly for the
correlation functions but not for the order parameters. One can first solve
the system for five of the $\rho_{p_i}$'s, obtaining
$$\left\{\eqalign{& \rho_{p_1} = \eta_1 C^2 \rho_{p_5} \cr
& \rho_{p_2} = \omega_1 C \rho_{p_5}  \cr
& \rho_{p_3} = \gamma C V \rho_{p_5}  \cr
& \rho_{p_6} = \omega_2 V \rho_{p_5}  \cr
& \rho_{p_9} = \eta_2 V^2 \rho_{p_5}  \cr}\right. \quad, \eqno(9)$$
where
$$\omega_1 = \exp{\left(-{d - b \over z \tau}\right)}, \qquad
\omega_2 = \exp{\left(-{d + b \over z \tau}\right)}, \qquad
\gamma = \exp{-1 + \xi - 2d \over z \tau}, $$
$$\eta_1 = \exp{1 + \xi - 2(d - b) \over z \tau}, \qquad
\eta_2 = \exp{1 + \xi - 2(d + b) \over z \tau},$$
$$C = \left[{y_2 + y_1 \over 2(1 - y_2)}\right]^\alpha, \qquad
V = \left[{y_2 - y_1 \over 2(1 - y_2)}\right]^\alpha, \quad
{\rm and} \quad \alpha = {z - 1 \over z}.$$
Now $\rho_{p_5}$ can be determined by means of the condition
${\rm Tr}\rho_p = 1$, which gives $\rho_{p_5} = W^{-1}$, with
$$W = \eta_1 C^2 + \eta_2 V^2 + 2\gamma CV + 2\omega_1 C + 2\omega_2 V + 1.$$
Finally, solving for the $y_i$'s yields two coupled equations for the order
parameters $y_1$ and $y_2$
$$\left\{\eqalign{
& y_1 = {\omega_1 C - \omega_2 V + \eta_1 C^2 - \eta_2 V^2 \over W} \cr
& y_2 = {\omega_1 C + \omega_2 V + \eta_1 C^2 + \eta_2 V^2 + 2\gamma CV \over
W}
\cr}\right. \quad , \eqno(10)$$
and the expressions for the correlations functions
$$\eqalign{
& y_3 = {\eta_1 C^2 + \eta_2 V^2 - 2\gamma CV \over W} \cr
& y_4 = {\eta_1 C^2 - \eta_2 V^2 \over W} \cr
& y_5 = {\eta_1 C^2 + \eta_2 V^2 + 2\gamma CV \over W} \cr}\quad . \eqno(11)$$
\par
It can be easily checked that eqs. (10) are substantially equivalent to the
equations of
Kikuchi's Natural Iteration Method$^{[16]}$, and thus they can be solved by
iteration. Furthermore, they involve a reduced set of parameters, that is,
the order parameters $y_1$ and $y_2$, instead of the whole set of pair density
matrix elements $\rho_{p_i},$ $i = 1, \ldots 9$ contained in Kikuchi's
equations.
\medskip

{\bf 3. Stability and multicritical points} \rm
\smallskip
In this section we derive the equations for the lines of critical and
tricritical points, that will be useful in the next section to determine
the regions of the phase space where reentrance occurs. We begin with the
determination of the stability regions for the paramagnetic phase, that
will lead us to the equations for the lines of critical points. \par
Equations (10) and (11) are not sufficient to satisfy the $f$-minimum
requirement, because they have been derived from the stationarity
conditions (8). It is a well-known fact of analysis that $f$ has a minimum in
$\{y_i\}$ if and only if the Hessian matrix $H \equiv H(\{y_i\})$ with
elements $H_{ij} = \displaystyle {\partial^2 f \over \partial y_i \partial
y_j}$
is positive definite. In turn, $H$ is positive definite if and only if
$\det H_i > 0$, $i = 1, \ldots 5$, where
$$H_i = \pmatrix{H_{11} & \cdots & H_{1i} \cr
\vdots & \ddots & \vdots \cr
H_{i1} & \cdots & H_{i1} \cr}.$$\par
We have solved the inequalities $\det H_i > 0$, $i = 1, \ldots 5$ for the
paramagnetic (that is, $y_1 = y_4 = 0$) solution of (10)-(11) in absence of
magnetic field (from this section on, we will take $b = 0$,
and then $\omega_1 = \omega_2 = \omega$, $\eta_1 = \eta_2 = \eta$),
in order to determine the stability conditions for the paramagnetic
phase. The first three inequalities are identically satisfied because of
the constraints (7) on the density matrix elements. The fourth one gives the
condition $y_2 > (z - 1) y_3$. Solving the associated equation yields
$$e^{d/\tau} = 2 \zeta (\gamma_0 - 1)
\left[ {\zeta \gamma_0 \over \zeta (\gamma_0 - 1) + 1} \right]^{z - 1},
\eqno(12)$$
which is satisfied by the second order critical temperature, where
$$\zeta = e^{\xi / z \tau}\cosh {1 \over z \tau}, \qquad
\gamma_0 = (z - 1) \tanh {1 \over z \tau}.$$
Finally, the fifth inequality adds the condition
$$(z - 2)y_2^2 + y_2 - (z - 1) y_5 > 0, \eqno(13)$$
which is identically satisfied for $\xi < -1$. \par
Plotting, for a fixed value $\xi > -1$, the solutions to eq. (12) and to the
equation corresponding to (13), one obtains, if $z \ge 3$
 (if $z = 2$ one has no phase transition), the diagram shown in Fig. 1.
The phase space is thus naturally divided in five regions $S_i, i = 1,
\ldots 5$. Numerical analysis shows that
in the region $S_1$ no stable paramagnetic phase can exist, in
$S_2$, $S_3$ and $S_4$ there can exist only one stable paramagnetic phase,
and in $S_5$ two stable paramagnetic phases can exist$^{[6]}$,
one with $y_2 < 1/2$, that will be denoted by Para$_-$,
and another with $y_2 > 1/2$, that will be denoted by Para$_+$.
In this region there is a first-order Para$_-$-Para$_+$
transition surface ending at
the line of critical points $C^{[1]}$, at which $y_2 = 1/2$.
The behavior of the
order parameter $y_2$ around $C$ is shown in Fig. 2. \par
The equations of this critical line can be
determined analytically by imposing that the two solutions to the equation
associated with (13) coincide.
One thus obtains, choosing $\tau$,
normalized temperature, as independent variable
$$d = \tau\left[\ln 2 - z \ln \left(1 - {2 \over z}\right)\right], \qquad
\xi = z \tau \ln {\left(\displaystyle{z \over z - 2}\right)^2 \over
\cosh \displaystyle{1 \over z \tau}} \quad .\eqno(14)$$ \par
Within these scheme it is also possible to determine the equations for the
tricritical line, which separates second order transitions from first order
ones. Indeed eqs. (10) can be written, to the 2nd order in $y_1$, in the
form (the solution $y_1$ = 0 has been factored away)
$$\left\{\eqalign{& p_0 + p_2 y_1^2 = 0 \cr
& q_0 + q_2 y_1^2 = 0 \cr}\right. \quad , \eqno(15)$$
where
$$p_0 \equiv p_0(y_2) = 2(\eta + \gamma)V_0^2 + 4 \omega V_0 + 1 -
2 V_0 {\alpha \over y_2}(\omega + 2 \eta V_0),$$
$$q_0 \equiv q_0(y_2) = 2 (\eta + \gamma) y_2 V_0^2 + 4 \omega y_2 V_0 + y_2
- 2 (\eta + \gamma) V_0^2 - 2 \omega V_0,$$
$$\eqalign{& p_2 \equiv p_2(y_2) = \cr
& = 2 {V_0 \over y_2^2} \left[
\eta V_0 {2\alpha \choose 2} - \gamma V_0 \alpha + 2 \omega {\alpha \choose 2}
\right]
- 2 {V_0 \over y_2^3}\left[
\omega {\alpha \choose 3} + \eta V_0 {2\alpha \choose 3}\right], \cr}$$
$$\eqalign{& q_2 \equiv q_2(y_2) = \cr
& = 2 {V_0 \over y_2} \left[
\eta V_0 {2\alpha \choose 2} - \gamma V_0 \alpha + 2 \omega {\alpha \choose 2}
\right] - 2 {V_0 \over y_2^2}\left[
\omega {\alpha \choose 2} + \eta V_0 {2\alpha \choose 2}
- \gamma V_0 \alpha \right], \cr}$$
with $V_0 = \displaystyle{\left[y_2 \over 2(1 - y_2)\right]^\alpha}$.
Again, to the 2nd order in $y_1$, we have $y_2 = a + b y_1^2$ and (15) becomes
$$\left\{\eqalign{& p_0(a) + (p_2(a) + b p_0^\prime(a)) y_1^2 = 0 \cr
& q_0(a) + (q_2(a) + b q_0^\prime(a)) y_1^2 = 0 \cr}\right. \quad . \eqno(16)$$
Letting $y_1 = 0$ in (16) gives the second order critical temperature
again, while, requiring that (16) are identically satisfied, we have the
following system of equations for the tricritical line$^{[17]}$
$$\left\{\eqalign{& q_0 = 0 \cr
& p_0 = 0 \cr
& p_2 q_0^\prime - p_0^\prime q_2 = 0 \cr}\right. \quad , \eqno(17)$$
with all functions evaluated at $a$ (which, of course, must be determined
from one of the equations).
Eqs. (17) cannot be solved analytically for two of the three parameters
$\xi,d,\tau$, but one can solve them numerically at fixed $\xi$ by looking
on the second order critical line given by eq. (12) for the point at which
the last of eqs. (17) changes sign. The values of $\tau$ and $d$ so
obtained are plotted in Fig. 3a-b for typical values of the coordination
number. Our results are compared, for $z = 6$,
with the effective field theory ones by
Tucker$^{[17]}$ in Table I, showing that our temperatures are slightly
lower, with an agreement to within 5\%.\par
Finally, let's notice that in the limit $\tau \to 0$ the critical and
tricritical lines converge in $\xi = -1$, $d = 0$. \par

{\bf 4. Reentrant phenomenon} \par
\smallskip
At fixed $-1 < \xi < 0$, our calculations show that the phase diagram has the
general shape given in Fig. 4. The solid line represents a second order
transition, while the dotted one represents a first order transition
(determined numerically, by comparison of free energies).  We have
denoted by $d_t$, $d_0$ and $d_m$ respectively the value of $d$
at the tricritical point, at $\tau = 0$, and the maximum value of $d$ for which
a
transition can occur. It is evident from this diagram that we can have two
different kind of reentrances: when $d_0 < d < d_m$ we have, increasing the
temperature, a Para-Ferro transition followed by a Ferro-Para one (both
second order); when
$d_t < d < \min (d_0, d_m)$ we have a Ferro-Para-Ferro-Para sequence of
transitions, the first of which is first order, in contrast with the result
of Ref. 18, and the others second order. \par
We have already mentioned how $d_t$ can be determined in our scheme, while
$d_m$ can be easily evaluated numerically by means of eq. (12).
In order to determine $d_0$ we must minimize the zero temperature free energy
$$f_0 = {\cal H} = dy_2 - {1 \over 2}(y_3 + \xi y_5). \eqno(18)$$
This is a linear function of $(y_2,y_3,y_5)$ defined in a polyhedron,
and thus its minima must be located on the vertices of the polyhedron, which,
according to the constraints (7), are $P_0 \equiv (0,0,0)$, $P_{1/2} \equiv
(1/2,0,0)$ and $P_{1\pm} \equiv (1,\pm 1,1)$. The stable phase (that is,
the one with lowest free energy) can be paramagnetic only if $y_2 = 0$
because if $y_2 > 0$ values of $y_1$ different from zero would be allowed
and a little magnetic field would make the phase with $y_1 = 0$ unstable.
This means that the stable phase is paramagnetic if and only if the
absolute minimum of $f_0$ is located at $P_0$, that is, if
$d > d_0 = \displaystyle{1 + \xi \over 2}$. \par
The values of $d_0$, $d_t$ and $d_m$ for $z = 6$ are plotted in Fig. 5,
where the
region of the $(\xi,d)$ plane where the two kind of reentrances occur are
also indicated. For $z \le 4$ we have no reentrance in agreement with
Ref. 8 and in contrast with the results of Ref. 11, while for increasing
$z$ the reentrance regions become larger, as already noticed in Ref. 10.
The maximum $\xi$ for
which there exist Para-Ferro-Para reentrance are $\xi = -0.345, -0.283,
-0.242$  for $z = 6,8,12$ respectively. We also give, for the
Ferro-Para-Ferro-Para case, the order parameters, the correlation functions
(Fig. 6a-b) and the specific heat (Fig. 7). \par
If $\xi < -1$ there is no longer first order transition, because the
tricritical temperature approaches zero as $\xi$ tends to -1 (Fig. 3a).
Again for $d_0 < d < d_m$ we have a Para-Ferro-Para sequence of
transitions, but now $d_0$ has to be determined in a different way: indeed,
the transition near $\tau = 0$ is second order, and thus we have
$$d_0 = \lim_{\tau \to 0, \xi < -1} d_c(\tau) = 1 + \xi, \eqno(19)$$
where $d_c(\tau)$ is the second order critical value of $d$, obtained from
(12).
It must also be noticed that in this case the low temperature paramagnetic
phase might be a staggered quadrupolar phase$^{[10]}$. We do not give more
details about this case, because it would require to consider two different
sublattices, and thus to change the whole scheme. \par
If $0 < \xi < 3$ we have no reentrance because both the second order and
the first order transition lines have negative slope in the ($d,\tau$)
plane. \par
Let us examine the last case of reentrance. A numerical analysis shows that,
for fixed $\xi > 3$, the first
order Para-Ferro transition line in the $(d,\tau)$ plane starts from $\tau
= 0$, $d = d_0 = \displaystyle{1 + \xi \over 2}$ with positive
slope, while the second order one has negative slope, as reported in Fig.
8 (phase diagram for $z = 4$, $\xi = 3.3$). The point $P$ is a four phase
coexistence point$^{[6]}$,
meeting point of three first-order transition lines,
Para$_-$-Ferro, Para$_+$-Ferro and Para$_-$-Para$_+$, and the corresponding
value $d_p$ (that must be evaluated numerically) is the maximum value of $d$
for which a Para-Ferro transition can occur. For $d_0 < d < d_p$ we have a
Para$_-$-Ferro-Para$_+$ sequence of transitions, and thus another reentrance.
The values of $d_0$ and $d_p$ are plotted in Fig. 9.
We have checked that, even in this case, increasing the
coordination number of the lattice the reentrance region becomes larger.
Again we give plots of order parameters, correlation functions (Fig. 10a-b)
and specific heat (Fig. 11). \par
It is noteworthy that already the early mean field phase diagrams by
Blume, Emery and Griffiths$^{[1]}$ could suggest the existence of this kind
of reentrance. \par
Finally we remark that the four phase coexistence point $P$ and the
tricritical point $T$ (Fig. 8), which are well distinct in mean field
approximation and coincide in renormalization group
studies$^{[6]}$, are very close in our approximation. In
fact, their locations in the $(d,\tau)$ plane are
$P \equiv (2.17,0.75)$, $T \equiv (2.10,0.79)$ in mean field approximation,
and $P \equiv (2.16,0.59)$, $T \equiv (2.15,0.60)$ in our approximation. \par
\medskip

{\bf 5. Conclusions} \par
\smallskip
We have studied the Blume-Emery-Griffiths model in the pair approximation
of the cluster variation method, determining the phase space conditions for
the reentrant phenomenon. Our results show that there are three different
kind of reentrances, characterized by different transition sequences.
For negative values of the reduced quadrupole-quadrupole interaction
strength $\xi$ we have found that reentrance is allowed only for $z \ge 6$ and
for $\xi < \xi_m$, where $\xi_m$ depends on $z$. For positive values of
$\xi$ we have found a little region of the phase space where a new
Para$_-$-Ferro-Para$_+$ reentrance can occur. Both regions become larger
for increasing $z$, as already pointed out in the literature. \par
As by-products, we have given analytical expressions for the critical and
tricritical lines. \par
\vfill
\eject

\parindent 0pt
{\bf Table I.} Coordinates of the tricritical points in the $(d, z\tau)$
plane ($z = 6$). In the third column are reported results of Ref. 17 for
comparison. \par
\vskip .35 truein
\centerline{
\vbox{\offinterlineskip
\hrule
\halign{&\vrule#&
	\strut\quad\hfil#\hfil\quad\cr
height2pt&\omit&&\omit&&\omit&\cr
&$d$&&$z\tau$&&$z\tau_{eft}$&\cr
height2pt&\omit&&\omit&&\omit&\cr
\noalign{\hrule}
height2pt&\omit&&\omit&&\omit&\cr
&0.468&&1.61&&1.67&\cr
&0.506&&1.76&&1.85&\cr
&0.543&&1.92&&2.01&\cr
&0.580&&2.06&&2.17&\cr
&0.617&&2.19&&2.31&\cr
&0.654&&2.33&&2.45&\cr
height2pt&\omit&&\omit&&\omit&\cr}
\hrule}}
\vfill
\eject

\parindent 0pt

{\bf Figure captions} \par
\parindent .4 truein
\item{}
\itemitem{\hbox to .8 truein{Fig. 1 : \hfill}}
Stability regions for the paramagnetic phases at $z = 6$, $\xi = 3.3$.
\itemitem{\hbox to .8 truein{Fig. 2 : \hfill}}
Quadrupolar order parameter $y_2$ around the critical
point $C$ ($z = 6$, $\xi = 3,3$).
\itemitem{\hbox to .8 truein{Fig. 3a : \hfill}}
Values of $\tau$ vs. $\xi$ at the tricritical point for $z =
3,6,12$ (solid lines) and in mean field approximation (dashed line).
\itemitem{\hbox to .8 truein{Fig. 3b : \hfill}}
Values of $d$ vs. $\xi$ at the tricritical point for $z = 6$ (solid
line, changing $z$ the line does not vary significantly) and in mean
field approximation (dashed line).
\itemitem{\hbox to .8 truein{Fig. 4 : \hfill}}
Phase diagram for $z = 6$, $\xi = -0.9$. $T$ is the tricritical point.
The solid and
dashed lines represent respectively second and first order transitions.
\itemitem{\hbox to .8 truein{Fig. 5 : \hfill}}
Reentrance region at $-1 < \xi < 0$. Solid line = $d_0$,
dashed line = $d_m$, dot-dashed line = $d_t$. In the region $P$ the ground
state is paramagnetic and one has a PFP sequence of
transitions; in the region $F$ one has ferromagnetic ground state and
FPFP transition sequence.
\itemitem{\hbox to .8 truein{Fig. 6a : \hfill}}
Dipolar order parameter $y_1$ (solid line) and correlation function $y_4$
(dashed line), vs. $\tau$, for the FPFP reentrance
($z = 6$, $\xi = -0.5$, $d = 0.2475$).
\itemitem{\hbox to .8 truein{Fig. 6b : \hfill}}
Quadrupolar order parameter $y_2$ (solid line) and  correlation functions
$y_3$ (dashed line) and $y_5$ (dot-dashed line), vs. $\tau$,
for the case of Fig. 6a.
\itemitem{\hbox to .8 truein{Fig. 7 : \hfill}}
Specific heat vs. $\tau$ for the case of Fig. 6a.
\itemitem{\hbox to .8 truein{Fig. 8 : \hfill}}
Phase diagram for $z = 4$, $\xi = 3.3$. $T$ = tricritical
point, $C$ = critical point, $P$ = four phase coexistence point. The solid and
dashed lines represent respectively second and first order transitions.
\itemitem{\hbox to .8 truein{Fig. 9 : \hfill}}
Reentrance region at $\xi > 3$. Solid line =
$d_0$, dashed line = $d_p$.
\itemitem{\hbox to .8 truein{Fig. 10a : \hfill}}
Dipolar order parameter $y_1$ (solid line) and correlation function $y_4$
(dashed line), vs. $\tau$, for the P$_-$FP$_+$ reentrance
($\xi = 3.3$, $d = 2.16$).
\itemitem{\hbox to .8 truein{Fig. 10b : \hfill}}
Quadrupolar order parameter $y_2$ (solid line) and  correlation functions
$y_3$ (dashed line) and $y_5$ (dot-dashed line), vs. $\tau$,
for the case of Fig. 10a.
\itemitem{\hbox to .8 truein{Fig. 11 : \hfill}}
Specific heat vs. $\tau$ for the case of Fig. 10a.
\vfill
\eject

\parindent 0 pt
{\bf References}
\smallskip
\item{[1]} M. Blume, V.J. Emery and R.B. Griffiths, Phys. Rev. {\bf A4}
(1971) 1071.
\item{[2]} D. Mukamel and M. Blume, Phys. Rev. {\bf A10} (1974) 610.
\item{[3]} J. Sivardi\'ere, Phys. Rev. {\bf B6} (1972) 4284.
\item{[4]} D.M. Staul, M. Wortis and D. Stauffer, Phys. Rev. {\bf B9}
(1974) 4964.
\item{[5]} A.K. Jain and D.P. Landau, Phys. Rev. {\bf B22} (1980) 445.
\item{[6]} A.N. Berker and M. Wortis, Phys. Rev. {\bf B14} (1976) 4946.
\item{[7]} O.F. de Alcantara Bonfim and F.C. S\'a Barreto, Phys. Lett.
{\bf 109A} (1985) 341.
\item{[8]} F.C. S\'a Barreto, Rev. Bras. de Fisica {\bf 20} (1990) 152.
\item{[9]} K.G. Chakraborty, J. Phys. {\bf C21} (1988) 2911.
\item{[10]} T. Kaneyoshi and E.F. Sarmento, Physica {\bf A152} (1988) 343.
\item{[11]} T. Kaneyoshi, Physica {\bf A164} (1990) 730.
\item{[12]} C. Buzano and L. Evangelista, J. Magn. Magn. Mat.
\item{[13]} R. Kikuchi, Phys. Rev. {\bf 81} (1951) 988.
\item{[14]} G. An, J. Stat. Phys. {\bf 52} (1988) 727.
\item{[15]} H.A. Bethe, Proc. R. Soc. {\bf A150} (1935) 552.
\item{[16]} R. Kikuchi, J. Chem. Phys. {\bf 60} (1974) 1071.
\item{[17]} J.W. Tucker, J. Phys. Cond. Matt. {\bf 1} (1989) 485.
\item{[18]} O.F. de Alcantara Bonfim and C.H. Obcemea, Z. Phys.
{\bf B64} (1986) 469.
\vfill
\eject
\end